\documentclass[final,5p,times,twocolumn]{elsarticle}

\usepackage{hyperref,graphicx}
\usepackage{amsmath}
\usepackage{verbatim}
\usepackage{pdfpages}
\usepackage{float}
\usepackage{subcaption}
\usepackage[ruled,lined,longend]{algorithm2e}
\usepackage{array}
\usepackage{tabularx}

\makeatletter
\def\ps@pprintTitle{%
 \let\@oddhead\@empty
 \let\@evenhead\@empty
 \def\@oddfoot{}%
 \let\@evenfoot\@oddfoot}
\makeatother

\journal{}

\begin{document}

\begin{frontmatter}

\title{Blockchain-based Covid Vaccination Registration and Monitoring}
\author[diu]{Shirajus Salekin Nabil}

\author[diu]{Md. Sabbir Alam Pran}

\author[diu]{Ali Abrar Al Haque}

\author[diu]{Narayan Ranjan Chakraborty}

\author[laTrobe]{Mohammad Jabed Morshed Chowdhury\corref{cor1}}
\ead{M.Chowdhury@latrobe.edu.au}

\author[brac]{Md Sadek Ferdous}

\address[diu]{Daffodil International University, Dhaka 1207, Bangladesh}
\address[laTrobe]{La Trobe University, VIC, 3086, Australia}
\address[brac]{BRAC University, Dhaka 1212, Bangladesh}

\cortext[cor1]{Corresponding author}

\begin{abstract}
Covid-19 (SARS-CoV-2) has changed almost all the aspects of our living. Governments around the world have imposed lockdown to slow down the transmissions. In the meantime, researchers worked hard to find the vaccine. Fortunately, we have found the vaccine, in fact a good number of them. However, managing the testing and vaccination process of the total population is a mammoth job. There are multiple government and private sector organisations that are working together to ensure proper testing and vaccination. However, there is always delay or data silo problems in multi-organisational works. Therefore, streamlining this process is vital to improve the efficiency and save more lives. It is already proved that technology has a significant impact on the health sector, including blockchain. Blockchain provides a distributed system along with greater privacy, transparency and authenticity. In this article, we have presented a blockchain-based system that seamlessly integrates testing and vaccination system, allowing the system to be transparent. The instant verification of any tamper-proof result and a transparent and efficient vaccination system have been exhibited and implemented in the research. We have also implemented the system as "Digital Vaccine Passport" (DVP) and analysed its performance.

\end{abstract}


\end{frontmatter}


\section{Introduction}


Information technology (IT) is playing a vital role in the fight against covid-19. It helps to evaluate the outbreak of the covid-19 pandemic, coronavirus statistical breakdown, identifying covid-19 through various symptoms, vaccine advancement \cite{Ng202004}. It is heavily used for contract tracing around the world. However, these systems are designed and maintained independently. Therefore, they cannot communicate with each other. It is challenging for policymakers to get a consolidated view of transmission, testing, and vaccination. In addition, trust in the testing data and vaccination data was in question, especially in developing countries. For instance, there was a case of test fraud happened in Bangladesh. Hospitals in Bangladesh named Regent hospital, JKG healthcare, and Shahabuddin hospital were caught scamming people by creating fake covid tests, wrong treatment and a series of other irregularities. These cases raised serious trust issues amongst people both inside and outside the country \cite{shahed}\cite{shahbuddin}\cite{sabrina}.

Blockchain is a distributed ledger technology, which can address the limitations of the current mechanism. It can help to integrate multiple systems and at the same time allows all the parties to interact with the plans without interfering with others. All the parties will maintain the system, and the system will automatically update if there is any activity in the system. However, it will automatically stop any unlawful activity. It provides a transparent view of the data to all the parties, and it helps build trust. It also prevents corruption as nobody can manipulate the data. If anybody provides wrong input purposefully, it can be detected using traditional audit or re-testing. The bad actor can be easily identified and held responsible as they cannot change the record in the system due to the immutability of the blockchain.

Blockchain contributes vastly to the healthcare sector. A blockchain-based medical record system was implemented to help patients keep their logs more securely. A distributed ledger like blockchain helps those data to be kept private and safe. Already Blockchain proved its capability in other health-based research. Blockchain performed amazingly in recording personal health data where the data privacy, flexibility, and authenticity were ensured \cite{kung} \cite{ekblaw2016} \cite{chowdhury2019trust}.  Blockchain has been useful in different aspects of the management of covid-19 pandemic, and thus, a risk notification system and location and bluetooth based contact tracing system had been implemented to ensure tamper-free services \cite{DBLP:journals/corr/abs-2007-10529}. Furthermore the Chinese University of Hong Kong has proposed a concept, describing the structure of a blockchain based vaccine passport with health records \cite{vaccinepassportfulll}. Blockchain can also assure the safety, security, transparency and traceability for distributing covid vaccine \cite{rotbi2021blockchain}. Double layer Blockchain has been used for recording vaccine production and information also. Using a timestamp, the information of enterprises and vaccines becomes tamper-proof, and the validity period of the vaccine is measured. A data cutting system has been introduced for reducing space \cite{9107255}. In line with these works, this article presents an integrated blockchain-enabled testing and vaccination system. The core contributions of the articles are as follows:
\begin{itemize}
    \item Designing a blockchain-based system that can seamlessly integrate testing and vaccination systems.
    \item Implementation of a QR-based "Digital Vaccine Passport" (DVP) mechanism which will reduce the corruption in covid testing and vaccination.
\end{itemize}

We have discussed the the background knowledge in section \ref{sec:bacgroun}. System requirements and design are discussed  in sections \ref{sec:systemrequirement} and \ref{sec:systemdesign} respectively. In section \ref{sec:implementation}, we have discussed about the implementation. Performance evaluation is presented in section \ref{sec:performanceevaluation}. We have concluded the article with future research directions in section \ref{sec:conclusion}.

\section{Background \& Related Work}
\label{sec:bacgroun}
As mentioned, there are two different aspects of our proposed model: the testing and the vaccination. We will discuss both aspects separately in the next two subsections. 
\subsection{Covid Testing and Reporting Systems }
Several works have been done with immunity passports, covid certifications, covid testings, ``Digital Health Passport" (DHP) based on blockchain \cite{tsoi2021way}. All of these terms are the same, but they have been named differently. Those have been discussed below. Already a prototype app has been developed where users testing reports will be generated and transparency will be achieved as it is developed under blockchain technology \cite{9105054}. Benchmarking results have also been shown there. Qr code based verification system developed where users' data become hidden. 
The ICT division of Bangladesh Government introduced a centralized app called "SUROKKHA" \cite{surokkha}. Using that Bangladesh government is currently running the vaccination process.

DHP is based on the secured distributed network (blockchain) that works as a health passport \cite{angelopoulos2020dhp}. DHP can be used for travelling, work and various places and helps regain a country’s economy. It proves that the person is not affected or the person already has immunity from covid-19. DHP is a proactive measure that helps to prevent the virus from spreading more.  Works have been done on "Digital Contract Tracing" (DCT) also which carries both benefits and some limitations \cite{xu2020beeptrace}. A blockchain-based covid-19 vaccination passport, offering the vaccination status of an individual where the user’s identity has been ensured by retina scanning \cite{chaudhari2020framework}. Masesk’s Algorithm has been used for generating greyscale eye images. One’s vaccine status is updated by taking his/her biometric data and the vaccination info. The encrypted version of the bio-metric data has been stored in the blockchain platform. Both bio-metric information and blockchain provide higher security and scalability.  An immunity testing certification based on blockchain has also been proposed where the user’s data has been registered on a  blockchain platform regulated by the government \cite{bansal2020optimizing}. Hospitals are also there for testing purposes. User’s bio-metric data have been gathered, which enriches security.  Contact tracing has also been implemented by collecting user's phone number, geolocation and timestamp.  An app has also been developed to store user’s testing information on the blockchain platform, which offers trustworthiness and artificial intelligence for tracking patient’s location details and self-testing \cite{mashamba2020blockchain}.

\subsection{Vaccination Prioritisation}
It is not possible to vaccinate the whole world over night, even in a month, as vaccine production is limited, a prioritisation-based vaccination can ease the way of vaccination to ensure a fair vaccination. Even in countries like Bangladesh, it is a crying need. To set priorities at first, we should consider the prioritising criteria.

In the case of a flu outbreak, how to distribute finite vaccination supply is now being debated \cite{influenza}. Traditional vaccination tactics target individuals most at risk for serious consequences, such as seniors, but they overlook (1) the distinctive pandemic pattern of mortality risk migrating to younger ages, and (2) the projected poorer vaccine effectiveness in elderly, and (3) variations in the number of years left to live as a function of age. George Washington University combined these factors to predict  the number of life years lost (YLL)  at a specific age and the number of life years saved in a future pandemic based on the mortality patterns of the three historical pandemics, vaccine efficacy by age and the structure of the US population in 2000. A group of analysts utilised a numerical demonstration to compare five age-stratified prioritisation methodologies for inoculation \cite{bubar2021model}.

As reported by a study of MIT, calls for disposing of prioritisation for SARS-CoV-2 immunisation are developing in the midst of concerns that prioritisation decreases inoculation speed \cite{priotizationvaccination}. They utilised an "SEIR" model to consider the impacts of inoculation dispersion on open well-being, comparing the prioritisation approach and speed beneath moderation measures that are either facilitated amid the antibody roll-out or maintained through the conclusion of the widespread period. Another model called "SIDARTHE" was combined for estimating the spread of Covid-19 along with a data-based model which demonstrates the new cases of fatality rate and the cost of a healthcare system based on studying the Italian case \cite{giordano2021modeling}.

According to the US National Academy of Medicine (NAM) three groups have been emphasised to vaccinate first \cite{jama2020}. At first, the front line covid-19 health workers like Doctors, Nurses, and so on. Secondly, the people working in the sectors like education with in-person attendance, food supply, childcare, etc., are at high risk for covid-19. Finally, the third category is those people who are already sick from severe health condition. Prioritising the third category has been recommended by both UN and NAM. University of California has evaluated the ideal allotment of a limited immunisation supply within the United States over bunches separated by age and fundamental labourer status, which compels openings for social separation \cite{buckner2021dynamic}. In any case, based on the objective, more youthful fundamental labourers are prioritised to control spread or seniors to straightforwardly control mortality. The prioritisation also done based on age and the occupation where there exists a high infection rate \cite{Babus2020.07.22.20160143}. However, age was emphasised more than occupation. For occupation, the people whose presence in work contributes higher in GDP got more priority.

A Symptom prediction model based on AI has been used to classify symptoms under 12 classes \cite{Jamshidi2021}. Those classes have been set by analysing health data of 74 hospitals in Tehran, Iran. Among KNN, ANN, LDN, Random Forest, and Naive Bayes, Random forest achieved a good ROC and AUC weighted mean.






In our approach, we have developed an algorithm in blockchain that can set priorities. However, we have emphasised the areas which have a higher positive ratio according to the test result, as such areas are at high risk. After testing every user of our system, authority’s next work is to set the priorities. By pressing a single button, the system will calculate the ratio and sort the list in descending order according to the ratio.
Then the vaccination will start according to that ratio. The system will not allow the inoculator to vaccinate people from fewer priority areas. If he does so, he will get caught because there will be a miss-match between the vaccine storage data in our system and the available vaccines in the storage. Because of using blockchain here, systems data alteration is not possible. As the most vaccines have multiple doses, we have developed the system to allow the area with a lower ratio to be vaccinated only when the first dose is completed in the area with a higher ratio.

\subsection{Blockchain}
Blockchain is a decentralised system and very useful for privacy and transparency. It is a tamper-proof digital ledger technology used to find the solution to real-life issues \cite{chowdhury2019comparative}. Since its invention, blockchain has been utilised in many fields as a smart contract and bitcoin transaction system. Blockchain allows monetary functions to be performed in a distributed way through some cryptocurrencies such as Bitcoin, Ethereum, Litecoin, Monero and Zerocash \cite{FENG201945}. 
Blockchain have brought the concept of decentralisation where there will not be any central authority to take control over a system. While transactions happen these transaction records are received by the nodes of the network. Those will not be added to the network as a block until the acceptance of existing nodes of the network which is called verification. For adding a block in the network, a consensus needs to be achieved among the validators. 

\begin{figure}[htbp]
    \centering
    \includegraphics[width=8cm]{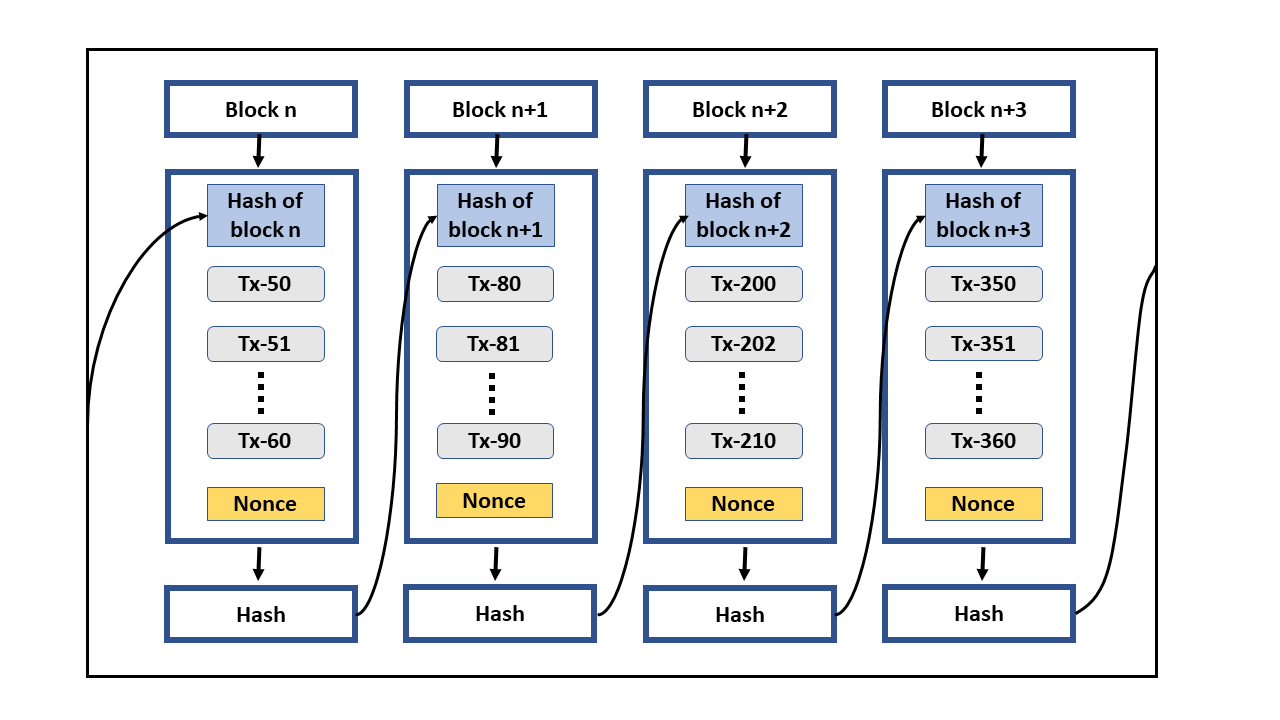}
    \caption{Proof of Work.}
    \label{usecase}
\end{figure}

In Bitcoin, a novel consensus algorithm called \textit{Proof of Work (PoW)} has been introduced \cite{nakamoto2008peer}. In this algorithm, the validators are called miners who are responsible for creating a valid block of transactions according to predetermined prerequisites. The block contains the hash of the previous block which has already been added to the chain and verified by the nodes Figure \ref{usecase}. POW implicitly defines the capability needed for a miner to add a block in the chain. Miners compete with each other to find a 32 bit number known as nonce. For example, to add a block miners need to be find a 32 bit number in which there will be a certain numbers of zero as the first numbers which change according to the network configuration. However, one can not make that number by force. To solve this miners have to use lots of computational power to find the number in first. Among all, the first miner who solves the puzzle and broadcasts the block in the network gets reward for their computational efforts. By that time the others who have been trying to solve that puzzle, cancel their mining process and take the hash of that block and try to solve the next block. After competing with this much of computational efforts it is assured that the miners did not cheat. 

\section{System Requirements}
\label{sec:systemrequirement}


Many organisations these days ask their employees to show positive covid-19 test certificates to get back into the office. Furthermore, many countries have restricted their borders from foreigners unless they produce the evidence that they are vaccinated. Besides, the whole vaccination process might create chaos since there are many candidates but not enough vaccines to inoculate. On top of that we have seen many cases of fraud covid-19 tests and unauthentic certification, which causes trust issues. Therefore, we need a proper and efficient system to bring back the trust in authority.

Considering the above mentioned issues, the proposed blockchain-based system need to meet the requirements stated below:

\begin{figure*}[h]
    \centering
    \includegraphics[width=13cm]{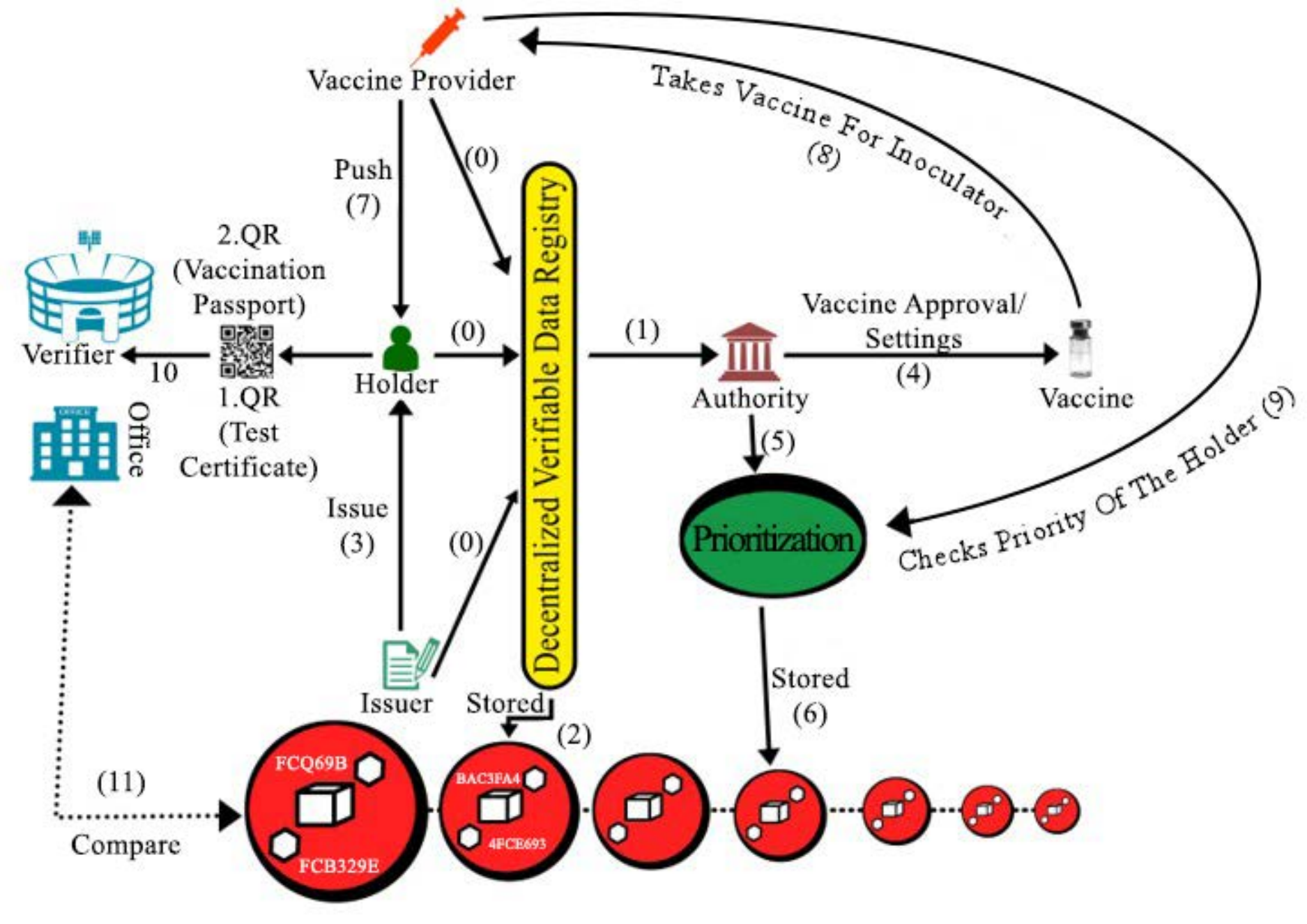}
    \caption{Workflow and use-case of covid19 test certification and vaccination model.}
    \label{workflow}
\end{figure*}

\begin{enumerate}
    \item \textbf{Fairness and Transparency:}  The vaccine distribution should be prioritised based on disease severity among the areas. The area at the top of the red zone list (where the ratio of positive and test cases is higher) must be inoculated first. The candidates who are tested covid-19 negative will get the higher priority. In this way, the vaccination process can be done without any chaos with transparency and fairness.To achieve that the system should store all the data and can automatically prioritise the area and the cohort. 

    \item \textbf{Battle Corruption:} In a traditional IT (Information Technology) system, it is always possible to alter the original data if the authority wants. Since it is centralised, the authority has the power over the system. Therefore, we need a decentralised or distributed system which cannot be changed or manipulated by one single authority.  
    
    \item \textbf{Seamless Integration:}  In multi-organisational systems, it always causes delay or data silo decreasing the network’s throughput, but in our blockchain-based proposed model, both testing and vaccination system and the prioritisation for vaccination are seamlessly integrated for transparency and user flexibility. We need a single platform which can seamlessly talk/communicate with multiple systems.
    
    
    \item \textbf{Cyber Attack Resilient:} The system should be designed in a way that it can withstand any cybersecurity attack.

\end{enumerate}

\subsection{Process in the System}
Figure \ref{workflow} gives a visual representation of our model. We have implemented a web-based blockchain application where all the operations will be performed.

\begin{figure*}[h]
    \centering
    \includegraphics[width=13cm]{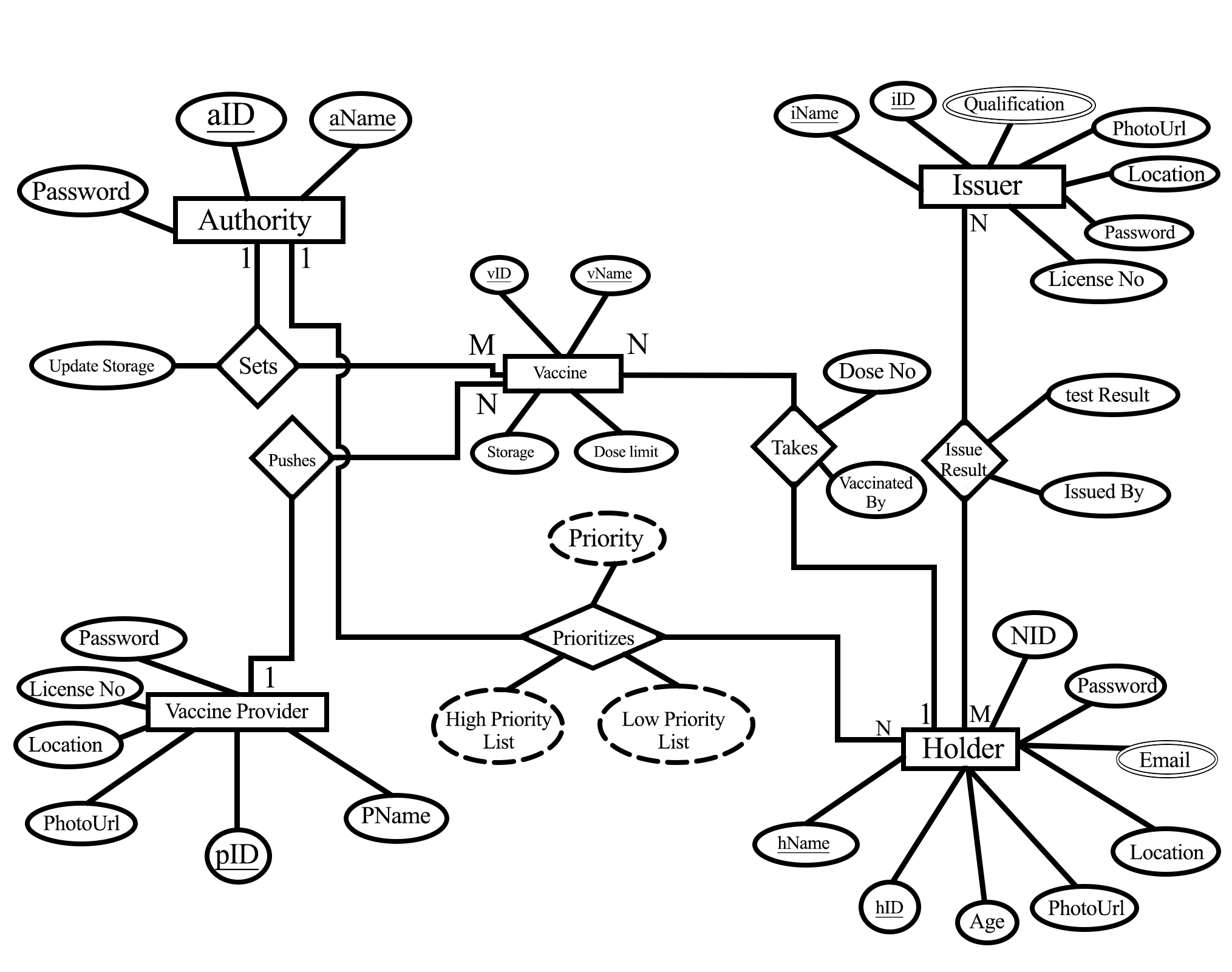}
    \caption{ER Diagram.}
    \label{fig:ER}
\end{figure*}

\section{System Design}
\label{sec:systemdesign}
In this research, our main concern is presenting a covid-19 testing certificate and prioritisation-based safe vaccination system followed by a vaccination passport for Bangladesh that has not been implemented yet. In our proposed system, there are five entities: ``Holder", ``Issuer", ``Vaccine Provider", ``Verifier", and ``Authority/Government". The activities of each entity are as discussed next.

\textbf{Authority/Government:}
Authority will be the person or organisation who governs the whole system. It has to bear the initial deployment cost of the project. Verified registration of each entity except the verifier will be ensured under the governance of the authority. It will accept or reject their sign up requests based on the NID (National Identity) of the holder and licence number of the issuer and vaccine provider(s). These two will be checked using the government's central database. Also, it will report to the system about new batches of vaccines as soon as it is received providing every detail of it.  Since the whole system aims to facilitate the holders or citizens of the country and ensure that no one’s fundamental rights are violated, proper distribution of vaccines should be established. There is no alternative to prioritise the candidates for vaccination. Authority will be playing this vital role of prioritisation by pressing a single button.

\textbf{Issuer:}
Issuers are the designated person who issue test results of covid19. After the covid-19 tests, test results will be provided to the holders by issuers in the form of QR code which will serve as a test certificate.

\textbf{Vaccine Provider:}
Vaccine providers are the hospitals who have the vaccine for mass inoculation. The priority list is strictly maintained under the system while the inoculation process is ongoing, so there will be no scope for venality. After the vaccination of each holder, the vaccine provider will provide them a "Digital Vaccine Passport" (DVP) in the form of a QR (Quick Response) code.

\textbf{Holder:}
Holders are the general people who will be inoculated. After the test and vaccination, each citizen will receive two QR codes, the test certificate and the digital vaccine passport, which will serve them as a gateway license to most restricted areas due to the pandemic. Holder can also control his/her profile privacy settings and see the information about priority list and vaccine storage for gaining the trust.

\textbf{Verifier:}
Verifiers are not required to register into the system, since they do not have any direct activities in the system. Offices, organisations, educational institutions, airports authority, etc. can play the role of verifiers to check a person’s eligibility for accessing the restricted areas due to pandemic. For doing so a verifier can simply scan the QR codes of the holder’s test certificate and digital vaccine passport and will confirm the necessary information about that holder.\\*
Since the entire process is maintained under a blockchain network, it can achieve high transparency and fairness.

\textbf{\textit{Prerequisite/ “On-boarding” step:}} The on-boarding is a three-step process as discussed next: 
\begin{itemize}
    \item \textbf{Step-0:} Every vaccine provider, issuer and holder have to make a request to sign up to the system as per their specific roles. It is a two-tesp processes as 
    \item \textbf{Step-1:} For identity authentication, their roles and professions are then verified by the governmental authority. Their sign up requests will be approved or rejected on the basis of authentication of their provided data. Providing false information may cause permanent banning if detected. 
    \item \textbf{Step-2:} Each user can access the system if and only if their sign up request is accepted and has an entry to the blockchain.
\end{itemize}



\begin{figure*}[h]
\label{fig:uml}
    \centering
    \includegraphics[width=12cm]{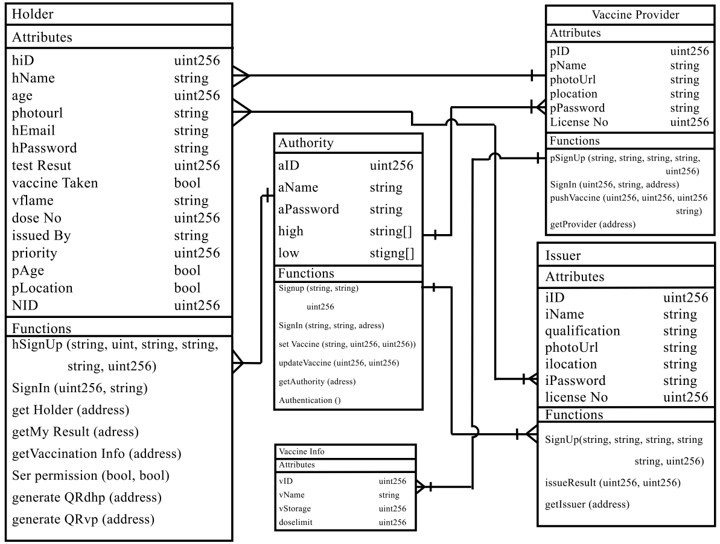}
    \caption{Class Diagram of the Proposed Blockchain-based System}
    \label{fig:uml}
\end{figure*}

\textbf{\textit{Issuing Covid Test Results:}} The end-users will provide their information like name, age, location and photo to register themselves for testing. Then they will apply for testing. The authorised issuers will provide the test result. In \textbf{Step-3}, the test result along with the issuer's ID will be generated as a QR code. This QR code can be used as a \textbf{Test Certificate} (QR1). 

\textbf{\textit{Vaccine Approval/Settings:}} The authority has access to store the vaccines data and update the vaccines storage only by positive margin if new batches are imported. This is regarded as \textbf{Step-4}.

\textbf{\textit{Prioritisation:}} As the vaccine production and distribution toward several countries are limited in quantity. So a monitored and authentic way is a crying need. At first, we have generated a location-wise ratio of positive cases among total tested cases. The higher rate locations are then selected as a higher priority as the affected cases and spreading probability are higher there. We know that vaccines are essential for those who do not have any immunity against the virus. So in \textbf{Step-5}, the negatively tested people are prioritised to mitigate the spreading of this deadly virus.  The list of the holders after prioritisation is stored in blockchain for a fair and transparent vaccination. This list will be checked at the time of each and every inoculation (\textbf{Step-6}).

\textbf{\textit{Inoculation:}} The inoculation is a three-step process as discussed next: 
\begin{itemize}
 \item \textbf{Step-7:} The vaccine provider is allowed for servicing the inoculation process and providing the vaccine passport.
 \item \textbf{Step-8:} Vaccine providers request the authority for vaccines to continue the inoculation process in. 
 \item \textbf{Step-9:} Vaccine providers will ask the holder for showing his/her test certificate and check whether the holder is eligible or not for the current inoculation wave according to the priority list.
 \end{itemize}
 After inoculation, the holder’s vaccination information along with the vaccine provider’s ID will be generated as a QR code. This QR code can be used as a \textbf{Vaccination Passport} (QR2). By doing these, the proper and efficient way of vaccination of those limited vaccines will be ensured.\\

\textbf{\textit{Verification:}} The verifier can be any organisation where it is necessary to ensure public safety. By this, a tamper-proof authentic result will be generated as these all are happening in the blockchain platform. Verifiers can extract the \textbf{QR1} code to verify any individual whether he/she is tested positive or negative \textbf{(Test Certificate)} and to verify any individual whether he/she has taken a vaccine, the \textbf{Vaccination Passport} or \textbf{QR2} code will be extracted in \textbf{step-10}. The information underneath the both QR codes are compared to the blockchain hashes for checking its validity as shown in \textbf{step-11} on Figure \ref{workflow}.



\subsection{Data Storage and Management}
Figure \ref{fig:ER} is the entity-relationship (ER) diagram of the proposed model, which represents the attributes that are stored for each entity (authority, issuer, vaccine provider, holder, vaccine) in their respective database. The hash of each entity’s name, id, and the hash of the priority list is stored in the blockchain since they are the primary key of the traditional database and must be kept securely.

\section{Implementation}
\label{sec:implementation}
In this section, we present the detailed implementation\footnote{Source Code: https://github.com/Salekin-Nabil/VDHP} of the proposed system. The system was implemented on top of Ethereum blockchain. The code was written in Solidy programming language on the Remix IDE, which is also used to compile and evaluate smart contracts. We have built three smart contracts so far; namely \textit{dhp} (Digital Health Passport), \textit{vaccination} and \textit{locationInfo}. Figure \ref{fig:uml} represents the class diagram of our project. The attributes of each structure are as shown in this figure and the methods are explained below.

The dhp smart contract and vaccination smart contract can be associated with each other and both can be associated with the locationInfo smart contract, but the locationInfo smart contract is fully independent. Firstly, the dhp smart contract is used for the \textbf{covid-19 test certificate}. It includes the information about the issuer and the holder. Secondly, the vaccination smart contract stands for the \textbf{vaccination passport} or a certification of vaccination which includes 3 structures namely: vaccine, authority and vaccine provider. Finally, the locationInfo smart contract is storing the location information to prioritise them according to the rate of covid-19 positive cases and utilise them as per needed (i.e. calculating total number of tests, total covid-19 positive cases, total vaccine receiving candidates in an area). The system has been deployed on Ethereum Rinkeby test network \cite{rinkeby}. The following parts go into the specifics of each smart contract.

\subsection{dhp: SignUp (all interfaces in general)} 
\begin{figure}[H]
    \centering
    \includegraphics[width=0.7\linewidth]{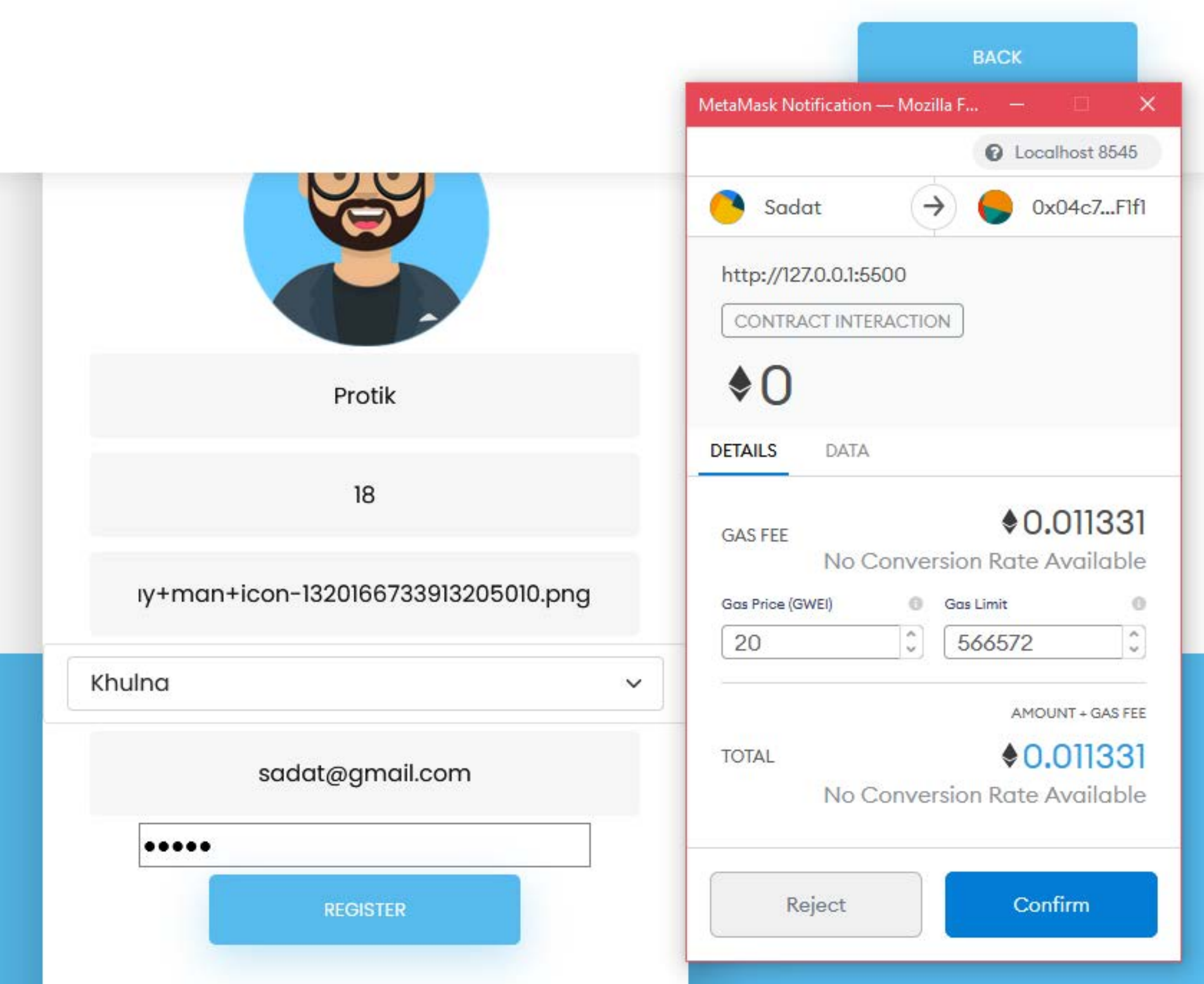}
    \caption{Sign Up.}
    \label{fig:signup}
\end{figure}

During the sign up process, an individual wallet owner can sign up for different roles (i.e. Authority, Issuer, Vaccine Provider, Holder) (Figure \ref{fig:signup}). 

To verify a wallet ,there are four checkers created with a view to identifying whether that wallet belongs to a role or not. It will also check whether the wallet is already registered for that specific role or not. After the initial checkers are satisfied, users can move forward to the main registration process by providing the necessary details. This function will then send an event to all of the involved entities, informing them of the change and the time. When interacting with the relevant entities, the blockchain client (BC) and gateway can use this event as part of its filtered events so that their records are modified appropriately. In future, we can also include captcha to improve the security of the login \cite{chowdhury2013captcha}.

\subsection{dhp: SignIn} 
During sign in, a new hash is generated combining the wallet address of the owner, the system ID (SID) of the user and password for verification. To verify the new hash, it is matched with all other combined hashes that already exist in the system regarding the specification of the user. A flag is used to determine whether the wallet is valid or not and redirected to the desired role’s menu accordingly in Algorithm \ref{signup}.

\begin{algorithm}[h]
    \caption{dhp:  SignIn}
    \label{signup}
     Initialisation of SID, Password, Address.
(Address holds the Ethereum Address of the function caller)\;
     Flag = “None”\;
   \uIf{new hash == existing Issuer hash}{%
     Flag = “issuer”\;}%
    \uElseIf{new hash == existing Holder hash}{%
    Flag = “holder”\;
    }
    \uElseIf{new hash == existing Authority hash}{%
    Flag = “authority”\;
    }
    \uElseIf{new hash == existing Vaccine Provider hash}{
    Flag = “vaccine provider”\;
    }
    \Else{%
    Show an error or failed to Sign In.\;
    }
\end{algorithm}

\subsection{ dhp: Issue Result (Issuer)} 
\begin{figure}[H]
    \centering
    \includegraphics[width=0.8\linewidth]{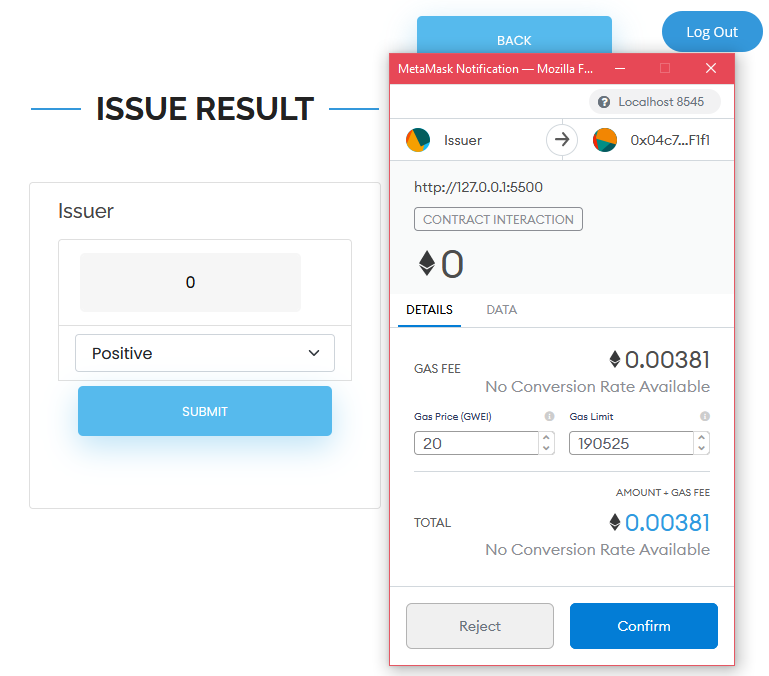}
    \caption{Test Result Issue.}
    \label{fig:issue}
\end{figure}

In case of issuing a test result the test will be done by the issuer first and then it has to be issued to the system Figure \ref{fig:issue}. In this algorithm, the owner of the issuer wallet is the only authorised entity that can issue test results to a holder. To verify a wallet that it belongs to a verified issuer, the address of the function caller would be checked to increase the security level. After the initial checkers are satisfied, the issuer can move forward to the insertion of the test result by providing the necessary details (holder ID, test result). This function will then send an event to all of the involved entities, informing them of the change and the time. 

When interacting with the relevant entities, the blockchain client (BC) and gateway can use this event as part of its filtered events so that their records are modified appropriately.


\subsection{Vaccination: Add Vaccine and Update Vaccine (Authority)}
\begin{figure}[H]
    \centering
    \includegraphics[width=0.8\linewidth]{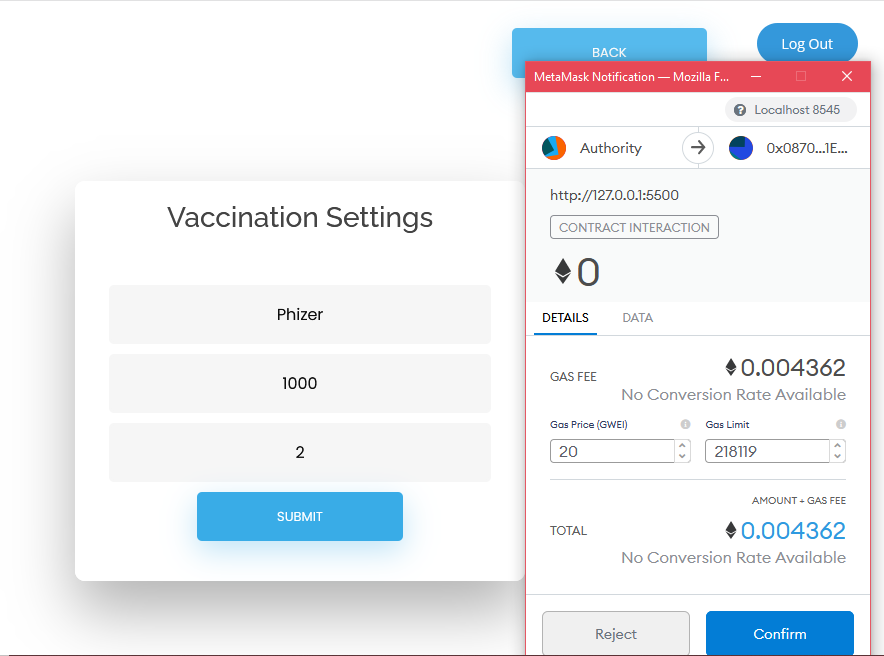}
    \caption{Add/Update vaccine.}
    \label{fig:vAdd}
\end{figure}

The user interface of adding vaccines to the approved list is shown in Figure \ref{fig:vAdd}. In this algorithm, the owner of the authority wallet is the only authorised entity that can add new vaccines to the system. To verify a wallet that it belongs to a verified authority the address of the function caller would be checked to increase the security level. After the initial checkers are satisfied, the authority can move forward to the insertion of the vaccines by providing the necessary details \textbf{(vaccine name, storage, dose limit)} and can also be updated in a similar way. 

This function will then send an event to all of the involved entities, informing them of the change and the time. When interacting with the relevant entities, the blockchain client (BC) and gateway can use this event as part of its filtered events so that their records are modified appropriately.




\subsection{Vaccination: Prioritisation (Authority)}  
Prioritisation can be done using Algorithm \ref{prioritisation}. In this algorithm, the owner of the authority wallet is the only authorised entity that can prioritise the holders for vaccination according to their location and test results (the negative holders of covid-19 who belong to the red zone area on the basis of positive case ratio per total number of tests). To verify a wallet that it belongs to a verified authority, the address of the function caller would be checked to increase the security level. After the initial checkers are satisfied, the authority can move forward to make a priority list of all the holders existing at that moment who have tested themselves with just a single click and everything else would be executed in the backend. 
\vspace{.5cm}

\begin{algorithm}[h]
\scriptsize
    \caption{ vaccination: Prioritisation - TX}
    \label{prioritisation}
     Initialisation of caller address.\;
     \textbf{Authority Checker()}\;
     j=8, k=0\;
     \While{j $<$ 16}{
     		\If{Totaltest $>$ 1}{
                    \While{traversing the entire holder’s length}{
                            \uIf{the holder has Positive result after testing Covid-19  $\And$$\And$  holder location == running location}{
                            Set the holder’s name to the low priority list\;
                            Increase the number that belongs to the running (j) priority\;
                            Holder Priority = j\;
                           }
                           \ElseIf{the holder has Negative result after testing Covid-19  $\And$$\And$ holder location == running location}{
                           Set the holder’s name to the high priority list\;
				Increase the number that belongs to the running (k) priority\;
				Holder Priority = k\;
                           }
                           j++\;
		               	   k++\;
                    }
            }
     }
\end{algorithm}

\vspace{.3cm}
Since we are prioritising the citizens of the eight divisions of Bangladesh and each division has two categories of holders (positive/negative), there are ($8x2$) $16$ levels of priority to prioritise the holders. The covid-19 negative holders who belong to the division which is at the top of the red zone list will be at the top priority for vaccination and they will be marked as the first level priority which indicates they should be vaccinated during the first wave of vaccination. Similarly the rest of the covid-19 negative holders will be marked by a level of priority from 2 to 8 according to the position of their division at the red zone list. The covid-19 positive holders will be marked by the levels of priority from 9 to 16 following the same pattern of the red zone list. This function will then send an event to all of the involved entities, informing them of the change and the time. When interacting with the relevant entities, the blockchain client (BC) and gateway can use this event as part of its filtered events so that their records are modified appropriately.


\subsection{Vaccination: Authentication and Approval of Issuer, Vaccine Provider and Holder Registration (Authority)}
Users can be added to the approved list of the respective role such as issuer, vaccine provider and holder registration using the following method. In this method, the owner of the authority wallet is the only authorised entity that can approve a user to enrol 
themselves to the system. To verify that a wallet belongs to a verified authority, the address of the function caller would be checked internally to increase the security level. After the initial checkers are satisfied, the authority can move forward to the authentication process with the help of the given details provided by the users while registering \textbf{(for vaccine providers and issuers - license number, for holders - NID)}. This function will then send an event to all of the involved entities, informing them of the change and the time. When interacting with the relevant entities, the blockchain client (BC) and gateway can use this event as part of its filtered events so that their records are modified appropriately.



\subsection{Vaccination: Inoculation/Push Vaccine (Vaccine Provider)}
\textbf{\textit{The First Dose:}} Inoculation process can be started in an unbiased way using Algorithm \ref{vaccine push}. In this algorithm, the owner of the vaccine provider wallet is the only authorised entity that can push vaccines to the holders. To verify that a wallet belongs to a verified authority, the address of the function caller would be checked to increase the security level. After the initial checkers are satisfied, the vaccine provider can move forward to the inoculation process. Lower priority holders can not be inoculated while higher priority holders are yet to inoculate. Due to the long time interval during each dose, lower priority holders can be inoculated after all the higher priority holders have finished taking their first doses. This function will then send an event to all of the involved entities, informing them of the change and the time. When interacting with the relevant entities, the blockchain client (BC) and gateway can use this event as part of its filtered events so that their records are modified appropriately.

\textit{\textbf{The Second Dose and Holder Elimination From The Priority List:}} Inoculation process can be successfully completed in an unbiased way using Algorithm \ref{vaccine push 2nd}. After completing the final dose the holder account will be removed from the vaccination priority list.

\subsection{dhp: Profile Permission (Holder)}

\begin{figure}[htp]
    \centering
    \includegraphics[width=0.8\linewidth]{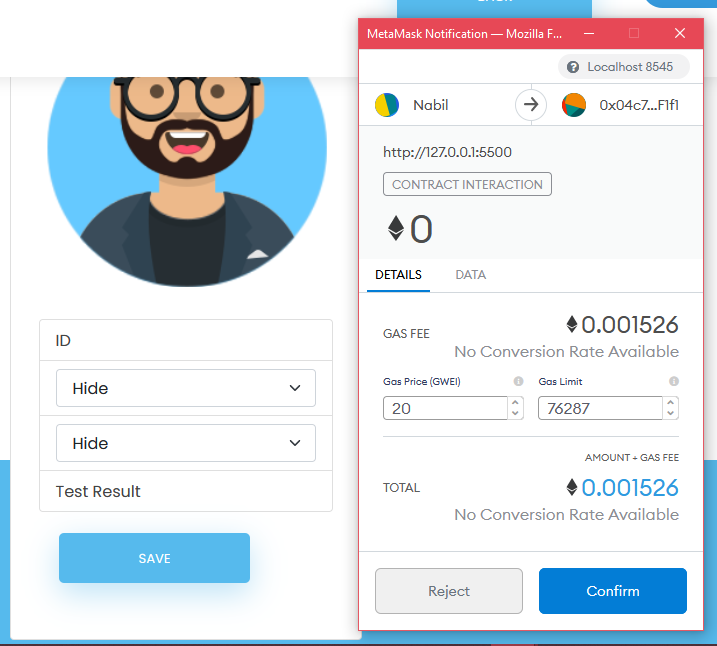}
    \caption{Permission Settings.}
    \label{fig:permission}
\end{figure}

A holder can set the permission of his profile about what to show and what not, the user interface of it is shown in figure-\ref{fig:permission}.
The owner of the Holder wallet is the only authorised entity that can change his/her profile permission. To verify that a wallet belongs to a verified holder, the address of the function caller would be checked to increase the security level. After the initial checkers are satisfied, the Holder can move forward to the permission setting process. This function will then send an event to all of the involved entities, informing them of the change and the time. When interacting with the relevant entities, the blockchain client (BC) and gateway can use this event as part of its filtered events so that their records are modified appropriately.

\subsection{dhp: Test Certificate (Holder)}
\begin{figure}[htp]
    \centering
    \includegraphics[width=0.8\linewidth]{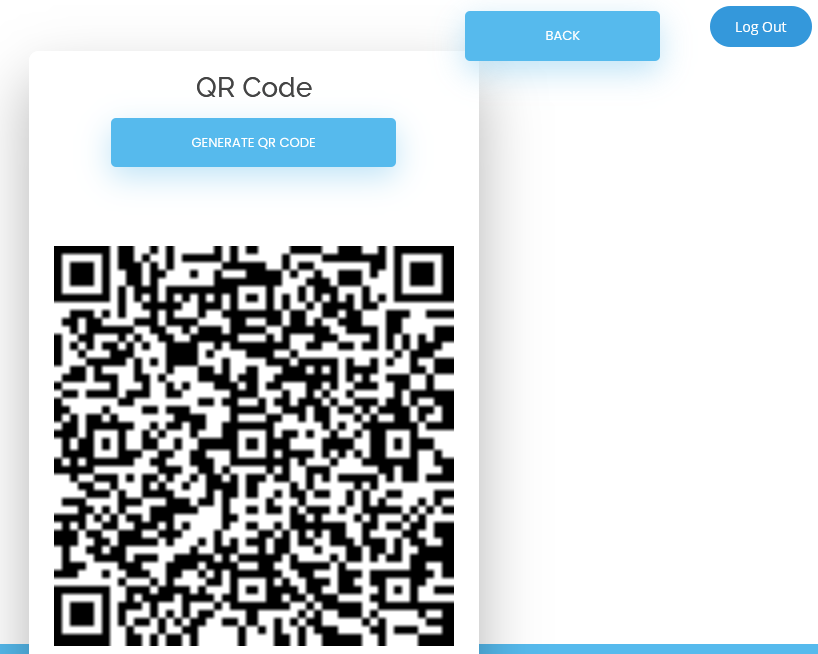}
    \caption{QR code of Test Certificate.}
    \label{fig:QR1}
\end{figure}

A holder can generate a QR code which has the information of his/her covid-19 test result (figure-\ref{fig:QR1}).
\vspace{0.3cm}

\begin{algorithm}[h]
\scriptsize
    \caption{ vaccination: Inoculation/Push Vaccine - TX
}
    \label{vaccine push}
   \textbf{Input:} Authority ID, Vaccine ID, Holder ID, Vaccine Name\;
    \textbf{Vaccine Provider Checker()}\;
    j=0, k=0\;
    \While{j $<$ 8}{
 			   \If{HighPriorityNo[j] != 0}{
               			[HighPriorityNo[j] - The number of holders in j-th priority who have n0t taken the first dose]\\
		break\;
               }
               j++\;
               }
               \If{j $<$ 8}{
               			\If{Holder belongs to the running priority}{
                        Push Vaccine;Vaccine Storage -=1\;
                        Dose Number (Holder) +=1\;
                        HighPriorityNo[j] -=1\;
                        
                        }
               }


   \ElseIf{j $\geq$ 8}{
   			\While{j $<$ 16}{
            		\If{LowPriorityNo[k] != 0}{
                    		break\;
                    }
                    j++; k++\;
                  
            }
            \If{ j $<$ 16}{
            		\If{Holder belongs to the running priority}{
                    	Push Vaccine;Vaccine Storage -=1\;
                        Dose Number (Holder) +=1\;
                        LowPriorityNo[j] -=1\;
                    }
            }
   }

\end{algorithm}


\begin{algorithm}[h]
\scriptsize
\caption{vaccination: Inoculation/Push Vaccine - TX}
\label{vaccine push 2nd}
\textbf{Input:} Authority ID, Vaccine ID, Holder ID, Vaccine Name{Continued from algorithm 3...}\;

\ElseIf {j $\ge$ 16}{
	k=0\; 
	\While {k $<$ 8}{
    \If {HighPriorityComp[k] != 0} {
[HighPriorityComp[k] - The number of holders in k-th priority who haven’t completed the dose]\\
			break\;}
		k++\;}

\If {k $<$ 8}{
	\If {Holder belongs to the running priority}{
    		Push Vaccine\;
			Vaccine Storage -=1\;
			Dose Number (Holder) +=1\;
			HighPriorityComp[k] -=1\;
			\If {Dose Number == Dose Limit} {Eliminate the Holder’s name from the vaccination list\;}
			}
		}
\ElseIf {k $\ge$ 8} {
		j=0\;
		\While {k $<$ 16}{
        	\If {LowPriorityComp[j] != 0} {[LowPriorityComp[j] - The number of holders in j-th priority who haven’t completed the dose]\\
				break\;}
                j++\;
				k++\;
		}
		\If {k $<$ 16}{
			If {Holder belongs to the running priority}{
            	Push Vaccine\;
				Vaccine Storage -=1\;
				Dose Number (Holder) +=1\;
				LowPriorityComp[k] -=1\;
					\If {Dose Number == Dose Limit}{Eliminate the Holder’s name from the vaccination list\;}
					}
			}
    }
 }

\end{algorithm}

%

\vspace{.2cm}
In this algorithm, the owner of the Holder wallet is the only authorised entity that generates his/her test certificate. To verify that a wallet belongs to a verified holder, the address of the function caller would be checked to increase the security level. After the initial checkers are satisfied, it generates a QR code containing necessary information \textbf{(holder’s name, age, photo, location, test result, the name of the issuer)} about the covid-19 test. Since it is just a query, it will not cost any ether.



\subsection{dhp: Vaccination Passport (Holder)}
A holder can generate a QR code which has the information of his/her covid-19 vaccination, similar to Figure \ref{fig:QR1}. In this algorithm, the owner of the Holder wallet is the only authorised entity that generates his/her vaccination passport. To verify that a wallet belongs to a verified holder the address of the function caller would be checked to increase the security level. After the initial checkers are satisfied, it generates a QR code containing necessary information \textbf{(holder’s name, vaccine taken, vaccine name, dose number, priority)} about the covid-19 Vaccination. Since it is just a query, it will not cost any ether.




\section{Performance Evolution}
\label{sec:performanceevaluation}
We benchmark our system using different parameters, such as latency, throughput, response time and failure rate. All the performance has been tested using Apache jMeter as it is comparatively the best load testing tool among loadrunner, microsoft visual studio (TFS), and siege \cite{jmeter, nevedrov2006using}.

We have run the simulations between 25 to 8500 users where each user has submitted a transaction (which is the registration of a holder under the Rinkeby test network) and recorded the response time, latency, throughput and failure rate. The bench markings were carried out in a PC with a Windows 10 64-bit operating system, having Intel Core-i5 8250U 1.80GHz CPU, 8GB DDR4 RAM, 1TB hard disk, 480GB SSD and Intel(R) UHD Graphics 620. 5 iterations of simulations were carried out and the average results are presented and discussed in the subsequent subsections.

\begin{figure*}[h]
\hspace*{\fill}
  \begin{subfigure}{0.49\textwidth}
    \includegraphics[width=0.8\linewidth]{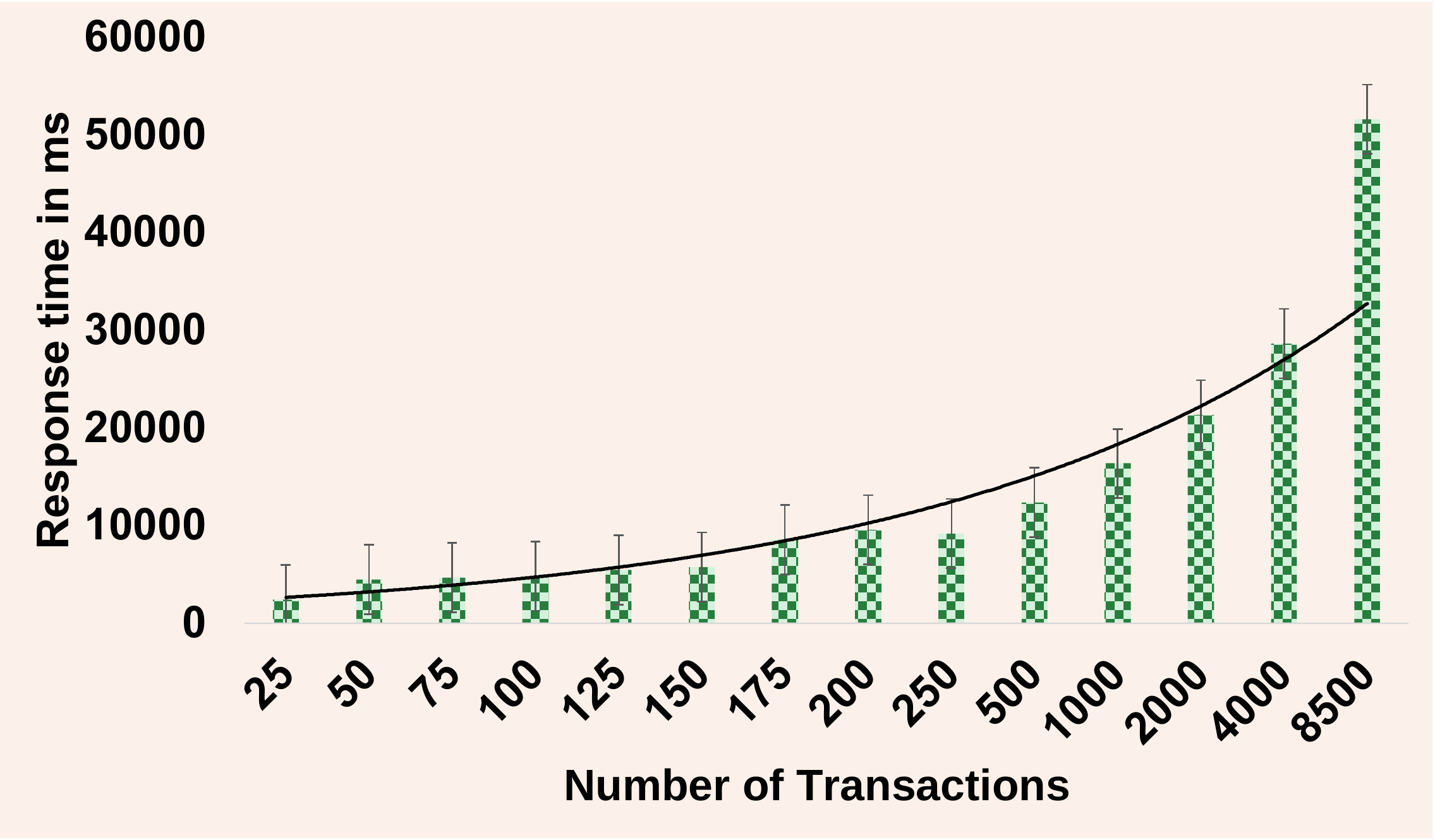}
    \caption{No. of Transactions vs Response Time} \label{Fig:RT}
  \end{subfigure}
  \begin{subfigure}{0.49\textwidth}
    \includegraphics[width=0.8\linewidth]{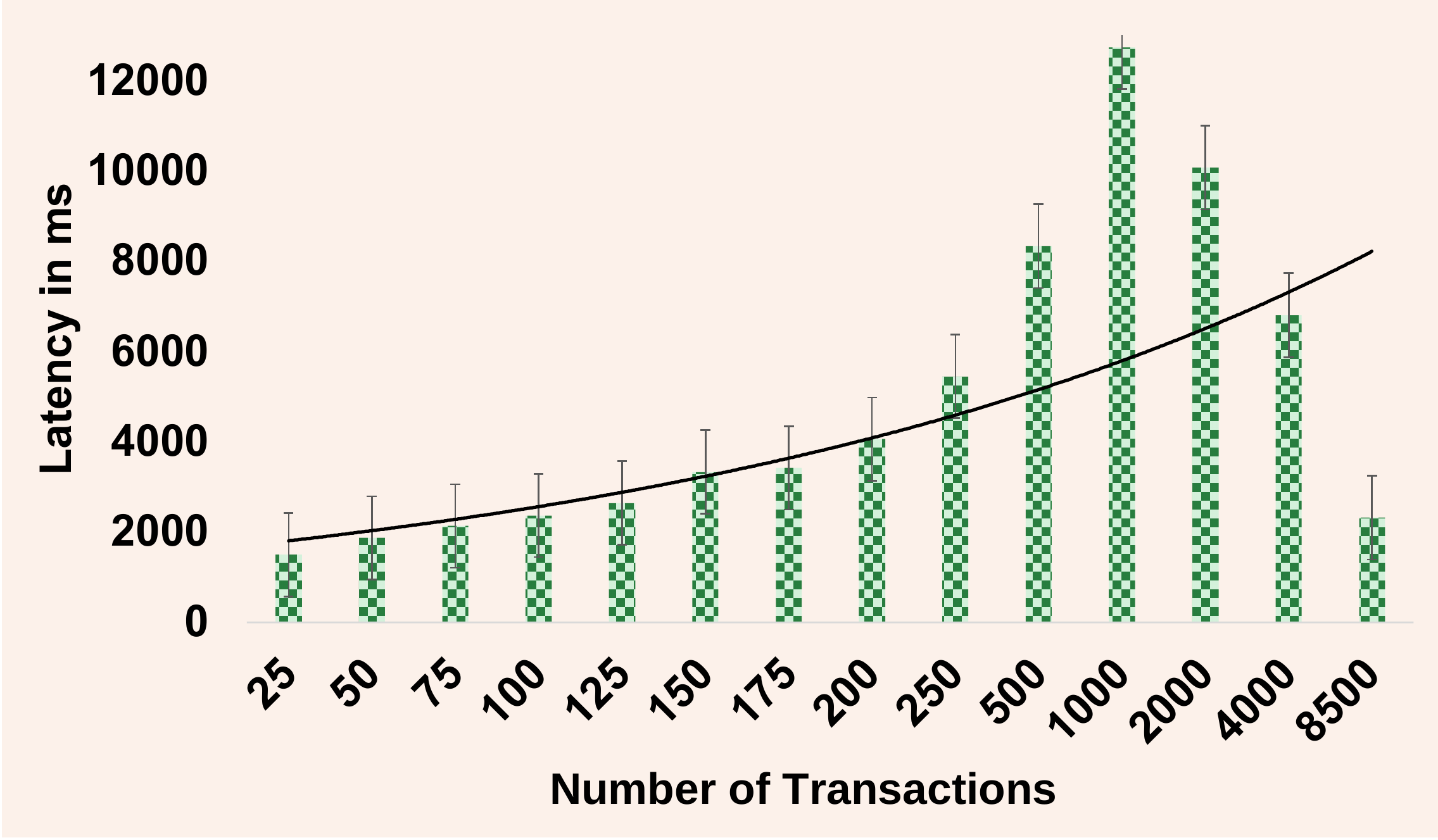}
    \caption{No. of Transactions vs Latency} \label{Fig:LT}
  \end{subfigure}%

  \hspace*{\fill} 
  \begin{subfigure}{0.49\textwidth}
    \includegraphics[width=0.8\linewidth]{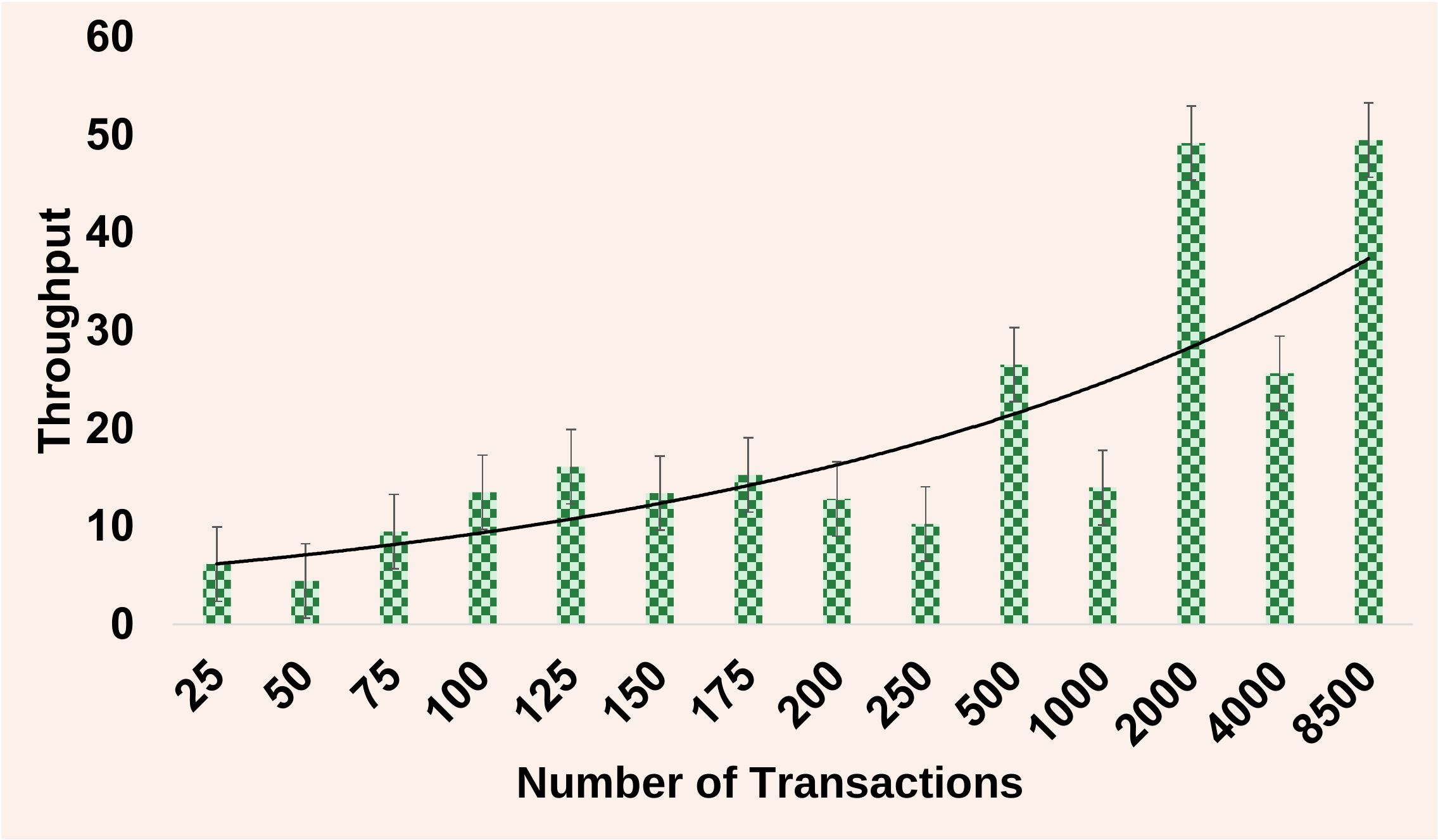}
    \caption{No. of Transactions vs Throughput}  \label{Fig:TH}
  \end{subfigure}
  \begin{subfigure}{0.49\textwidth}
    \includegraphics[width=0.8\linewidth]{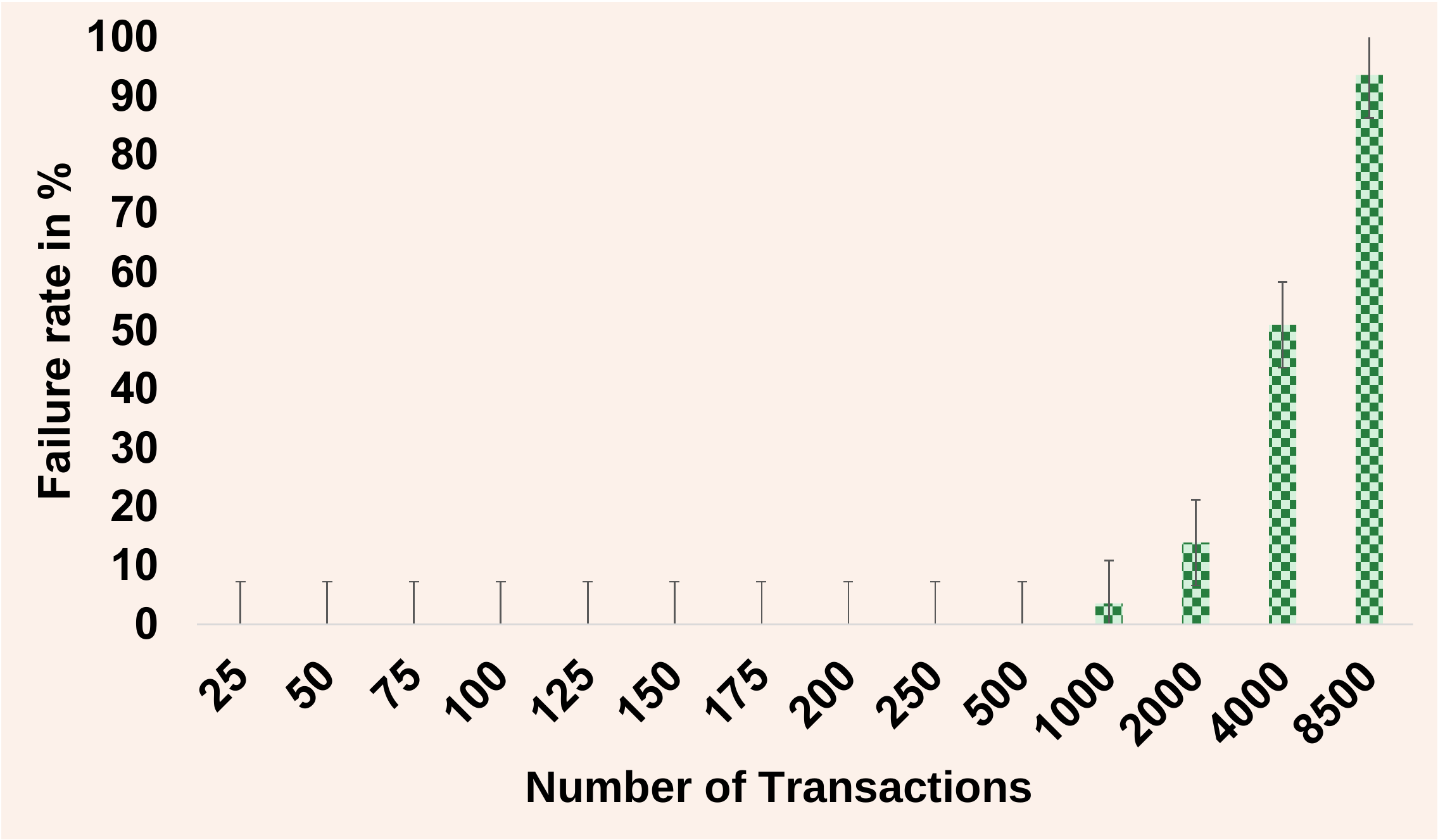}
    \caption{No. of Transactions vs Failure Rate}  \label{Fig:FR}
  \end{subfigure}
  \caption{Performance Analysis} \label{fig:tps}
\end{figure*}


\subsection{No. of Transactions vs Response Time}
The graph of the response time for the transaction is presented in Figure \ref{Fig:RT}. The response time starts to increase as the number of transactions increases. The increasing rate of the response time is almost exponential for comparing the diversity of response time with respect to the number of transactions, taking between 2412.96ms and 51653.7ms for 25 transactions and 8500 transactions respectively.

    
Thus, we can make an analysis of the system that response time is proportional to the number of transactions.
    
\subsection{No. of Transactions vs Latency}
The comparative latency for the transaction vs latency is presented in Figure \ref{Fig:LT}. The latency starts to increase as the number of transactions increases to 1000 transactions and then again starts to decrease from its peak value till the number of transactions reaches 8500.The increasing rate of the latency or the decreasing rate of it when it reaches the peak value is not as exponential as response time with respect to the number of transactions, taking between 1501.12ms and 12739.08ms for 25 transactions and 1000 transactions respectively and then falls back to 2320.732ms while reaching 8500 transactions .


\subsection{No. of Transactions vs Throughput}
We also studied the throughput for the transaction and it is presented in Figure \ref{Fig:TH}. The throughput shows an enormous diversity as the number of transactions increases which leads to a non-exponential analysis, transmitting between 6.17 tx/s and 49.51 tx/s for 25 transactions and 8500 transaction with a plenty of varieties in between.

    
\subsection{No. of Transactions vs Failure Rate}
The analysis of failure rate for the transaction is presented at Figure \ref{Fig:FR}. The failure rate starts to increase as the number of transactions increases. The increasing rate of the failure is almost exponential for comparing the diversity of failure rate with respect to the number of transactions, occurring between 0\% failure and 93.62\% failure for 25 transactions and 8500 transactions.

    
There occurs no failure with a small number of users (transactions) that is below 1000 and failure starts to occur when 1000 or more users hit for the transaction. Thus it is visible that failure rate is proportional to the number of users.

\subsection{Cost Analysis}
The deployment and transaction cost of all interfaces have been presented below in table \ref{tab}.

\begin{table}[h]
\caption{Gas Cost in USD on March 6, 2021}
\begin{center}
\begin{tabular}{|c|c|c|}
\hline
\textbf{Category} & \textbf{Gas Cost} & \textbf{Doller} \\
\hline
Contract Deploy(3)  & 0.210641  & 314.89 \\
\hline
Set Priority(300 people)  & 0.127718  & 190.93\\
\hline
Authority Sign Up  &0.00429  & 6.41\\
\hline
System Initialization  &0.0209  & 31.24\\
\hline
Update Storage  &0.001269  & 1.90 \\
\hline
Set Vaccine  &0.004362  & 6.52\\
\hline
Issuer Sign Up  & 0.008051  & 12.28\\
\hline
Issue Result  & 0.002292  & 3.50\\
\hline
Push Vaccine (2 Dose)  & 0.010489  & 16.00\\
\hline
Vaccine Provider Sign Up  & 0.007475  & 11.40\\
\hline
Holder Sign Up  &0.011652  & 17.78\\
\hline
Holder Permission  & 0.00147  & 2.24\\
\hline
\end{tabular}
\label{tab}
\end{center}
\end{table}


Due to the increment of ether cost the overall transaction become a great matter of fact. The price of the public crypto-currencies like bitcoin and ether are unpredictable. However this issue can be easily tackled by introducing private blockchain where the mining cost can be reduced. Hyperledger fabric will be a good alternative solution for that. 
  
\section{Conclusion}
\label{sec:conclusion}
It has been a year or more that the world is suffering with the invisible enemy sars-cov2 (covid-19). It is difficult to fight or create antibodies against viruses. In the past we have seen that the vaccine against viruses like influenza, ebola and so on took several years to achieve a suitable vaccine. Fortunately, due to the massive advancement in technology we have now lots of approved vaccines within just one year. However, due to production limitations, it is impossible to cover all people under vaccination within a very short period of time. So the possibility of chaos in vaccination raises. Countries like Bangladesh with huge populations need an authentic prioritisation based system, where proper vaccination will be assured without any chaos. Our implemented system offers all these criteria. Also authentic test report certification has been implemented where unbiased and counterfeit certificates have no chance to get rid from the authority. Recently, we have seen lots of scams recently related to false covid test certificates. Our blockchain-enabled deployed system mitigates the tampering possibilities and creates transparency. We have shown the cost efficiency and benchmark result where we proved that our system will provide all these services conveniently. Prioritisation offers the most unique but significant feature which will ensure optimised vaccination. With all these features, we believe that our system can be an effective tool to fight against covid-19.


\bibliography{mybibfile.bib}

\begin{thebibliography}{10}
\expandafter\ifx\csname url\endcsname\relax
  \def\url#1{\texttt{#1}}\fi
\expandafter\ifx\csname urlprefix\endcsname\relax\def\urlprefix{URL }\fi
\expandafter\ifx\csname href\endcsname\relax
  \def\href#1#2{#2} \def\path#1{#1}\fi

\bibitem{Ng202004}
P.~P. Nguyen~D.C., Dinh~M., S.~A., Blockchain and ai-based solutions to combat
  coronavirus (covid-19)-like epidemics: A survey., preprints202004\href
  {http://dx.doi.org/10.20944/preprints202004.0325.v1}
  {\path{doi:10.20944/preprints202004.0325.v1}}.

\bibitem{shahed}
H.~Sullivan, agencies, Global report: Bangladesh hospital owner accused of
  faking thousands of covid-19 test results\href
  {http://arxiv.org/abs/https://www.theguardian.com/world/2020/jul/16/
  global-report-bangladesh/hospital-owner-accused-of-
  faking-thousands-of-covid-19/test-results}
  {\path{arXiv:https://www.theguardian.com/world/2020/jul/16/
  global-report-bangladesh/hospital-owner-accused-of-
  faking-thousands-of-covid-19/test-results}}.

\bibitem{shahbuddin}
M.~S.~I. Tipu, A.~Alif, Covid-19 scam: Shahabuddin hospital md, 2 others
  remanded\href
  {http://arxiv.org/abs/https://www.dhakatribune.com/bangladesh/dhaka/2020/07/21/rab-hands-over-shahabuddin-hospital-md-to-police}
  {\path{arXiv:https://www.dhakatribune.com/bangladesh/dhaka/2020/07/21/rab-hands-over-shahabuddin-hospital-md-to-police}}.

\bibitem{sabrina}
M.~Saad, Sabrina washes her hands of jkg\href
  {http://arxiv.org/abs/https://www.thedailystar.net/frontpage/news/sabrina-washes-her-hands-jkg-1929997?fbclid=IwAR2c74VYFBtKLPjOXyDbakaVFv36yiYgrRqjNFMFUc48GuUB_4WWrI7r3ko}
  {\path{arXiv:https://www.thedailystar.net/frontpage/news/sabrina-washes-her-hands-jkg-1929997?fbclid=IwAR2c74VYFBtKLPjOXyDbakaVFv36yiYgrRqjNFMFUc48GuUB_4WWrI7r3ko}}.

\bibitem{kung}
H.~H. Kung, Y.-F. Cheng, H.-A. Lee, C.-Y. Hsu, Personal health record in fhir
  format based on blockchain architecture (2020) 1776--1788.

\bibitem{ekblaw2016}
A.~Ekblaw, A.~Azaria, J.~D. Halamka, A.~Lippman, A case study for blockchain in
  healthcare:“medrec” prototype for electronic health records and medical
  research data, in: Proceedings of IEEE open \& big data conference, Vol.~13,
  2016, p.~13.

\bibitem{chowdhury2019trust}
M.~J.~M. Chowdhury, M.~S. Ferdous, K.~Biswas, N.~Chowdhury, A.~Kayes,
  P.~Watters, A.~Ng, Trust modeling for blockchain-based wearable data market,
  in: 2019 IEEE International Conference on Cloud Computing Technology and
  Science (CloudCom), IEEE, 2019, pp. 411--417.

\bibitem{DBLP:journals/corr/abs-2007-10529}
J.~Song, T.~Gu, X.~Feng, Y.~Ge, P.~Mohapatra,
  \href{https://arxiv.org/abs/2007.10529}{Blockchain meets {COVID-19:} {A}
  framework for contact information sharing and risk notification system}, CoRR
  abs/2007.10529.
\newblock \href {http://arxiv.org/abs/2007.10529} {\path{arXiv:2007.10529}}.
\newline\urlprefix\url{https://arxiv.org/abs/2007.10529}

\bibitem{vaccinepassportfulll}
K.~K. Tsoi, J.~J. Sung, H.~W. Lee, K.~K. Yiu, H.~Fung, S.~Y. Wong, The way
  forward after covid-19 vaccination: vaccine passports with blockchain to
  protect personal privacy, BMJ Innovations 7~(2).

\bibitem{rotbi2021blockchain}
M.~F. Rotbi, S.~Motahhir, A.~E. Ghzizal, Blockchain technology for a safe and
  transparent covid-19 vaccination, arXiv preprint arXiv:2104.05428.

\bibitem{9107255}
S.~{Peng}, X.~{Hu}, J.~{Zhang}, X.~{Xie}, C.~{Long}, Z.~{Tian}, H.~{Jiang}, An
  efficient double-layer blockchain method for vaccine production supervision,
  IEEE Transactions on NanoBioscience 19~(3) (2020) 579--587.
\newblock \href {http://dx.doi.org/10.1109/TNB.2020.2999637}
  {\path{doi:10.1109/TNB.2020.2999637}}.

\bibitem{tsoi2021way}
K.~K. Tsoi, J.~J. Sung, H.~W. Lee, K.~K. Yiu, H.~Fung, S.~Y. Wong, The way
  forward after covid-19 vaccination: vaccine passports with blockchain to
  protect personal privacy, BMJ Innovations 7~(2).

\bibitem{9105054}
M.~{Eisenstadt}, M.~{Ramachandran}, N.~{Chowdhury}, A.~{Third}, J.~{Domingue},
  Covid-19 antibody test/vaccination certification: There's an app for that,
  IEEE Open Journal of Engineering in Medicine and Biology 1 (2020) 148--155.
\newblock \href {http://dx.doi.org/10.1109/OJEMB.2020.2999214}
  {\path{doi:10.1109/OJEMB.2020.2999214}}.

\bibitem{surokkha}
Covid-19 vaccine registration site now open\href
  {http://arxiv.org/abs/https://www.dhakatribune.com/bangladesh/2021/01/27/covid-19-vaccine-registration-app-site-now-open}
  {\path{arXiv:https://www.dhakatribune.com/bangladesh/2021/01/27/covid-19-vaccine-registration-app-site-now-open}}.

\bibitem{angelopoulos2020dhp}
C.~M. Angelopoulos, A.~Damianou, V.~Katos, Dhp framework: Digital health
  passports using blockchain--use case on international tourism during the
  covid-19 pandemic, arXiv preprint arXiv:2005.08922.

\bibitem{xu2020beeptrace}
H.~Xu, L.~Zhang, O.~Onireti, Y.~Fang, W.~J. Buchanan, M.~A. Imran, Beeptrace:
  blockchain-enabled privacy-preserving contact tracing for covid-19 pandemic
  and beyond, IEEE Internet of Things Journal.

\bibitem{chaudhari2020framework}
S.~Chaudhari, M.~Clear, H.~Tewari, Framework for a dlt based covid-19 passport,
  arXiv preprint arXiv:2008.01120.

\bibitem{bansal2020optimizing}
A.~Bansal, C.~Garg, R.~P. Padappayil, Optimizing the implementation of covid-19
  “immunity certificates” using blockchain, Journal of Medical Systems
  44~(9) (2020) 1--2.

\bibitem{mashamba2020blockchain}
T.~P. Mashamba-Thompson, E.~D. Crayton, Blockchain and artificial intelligence
  technology for novel coronavirus disease-19 self-testing (2020).

\bibitem{influenza}
M.~A. Miller, C.~Viboud, D.~R. Olson, R.~F. Grais, M.~A. Rabaa, L.~Simonsen,
  \href{https://doi.org/10.1086/589716}{{Prioritization of Influenza Pandemic
  Vaccination to Minimize Years of Life Lost}}, The Journal of Infectious
  Diseases 198~(3) (2008) 305--311.
\newblock \href
  {http://arxiv.org/abs/https://academic.oup.com/jid/article-pdf/198/3/305/18053670/198-3-305.pdf}
  {\path{arXiv:https://academic.oup.com/jid/article-pdf/198/3/305/18053670/198-3-305.pdf}},
  \href {http://dx.doi.org/10.1086/589716} {\path{doi:10.1086/589716}}.
\newline\urlprefix\url{https://doi.org/10.1086/589716}

\bibitem{bubar2021model}
K.~M. Bubar, K.~Reinholt, S.~M. Kissler, M.~Lipsitch, S.~Cobey, Y.~H. Grad,
  D.~B. Larremore, Model-informed covid-19 vaccine prioritization strategies by
  age and serostatus, Science 371~(6532) (2021) 916--921.

\bibitem{priotizationvaccination}
N.~Agarwal, A.~Komo, C.~A. Patel, P.~A. Pathak, M.~U. {\"U}nver, The trade-off
  between prioritization and vaccination speed depends on mitigation measures,
  Tech. rep., National Bureau of Economic Research (2021).

\bibitem{giordano2021modeling}
G.~Giordano, M.~Colaneri, A.~Di~Filippo, F.~Blanchini, P.~Bolzern,
  G.~De~Nicolao, P.~Sacchi, P.~Colaneri, R.~Bruno, Modeling vaccination
  rollouts, sars-cov-2 variants and the requirement for non-pharmaceutical
  interventions in italy, Nature Medicine 27~(6) (2021) 993--998.

\bibitem{jama2020}
G.~Persad, M.~E. Peek, E.~J. Emanuel,
  \href{https://doi.org/10.1001/jama.2020.18513}{{Fairly Prioritizing Groups
  for Access to COVID-19 Vaccines}}, JAMA 324~(16) (2020) 1601--1602.
\newblock \href
  {http://arxiv.org/abs/https://jamanetwork.com/journals/jama/articlepdf/2770684/jama\_persad\_2020\_vp\_200200\_1603319730.01653.pdf}
  {\path{arXiv:https://jamanetwork.com/journals/jama/articlepdf/2770684/jama\_persad\_2020\_vp\_200200\_1603319730.01653.pdf}},
  \href {http://dx.doi.org/10.1001/jama.2020.18513}
  {\path{doi:10.1001/jama.2020.18513}}.
\newline\urlprefix\url{https://doi.org/10.1001/jama.2020.18513}

\bibitem{buckner2021dynamic}
J.~H. Buckner, G.~Chowell, M.~R. Springborn, Dynamic prioritization of covid-19
  vaccines when social distancing is limited for essential workers, Proceedings
  of the National Academy of Sciences 118~(16).

\bibitem{Babus2020.07.22.20160143}
A.~Babus, S.~Das, S.~Lee,
  \href{https://www.medrxiv.org/content/early/2020/12/03/2020.07.22.20160143}{The
  optimal allocation of covid-19 vaccines}, medRxiv\href
  {http://arxiv.org/abs/https://www.medrxiv.org/content/early/2020/12/03/2020.07.22.20160143.full.pdf}
  {\path{arXiv:https://www.medrxiv.org/content/early/2020/12/03/2020.07.22.20160143.full.pdf}},
  \href {http://dx.doi.org/10.1101/2020.07.22.20160143}
  {\path{doi:10.1101/2020.07.22.20160143}}.
\newline\urlprefix\url{https://www.medrxiv.org/content/early/2020/12/03/2020.07.22.20160143}

\bibitem{Jamshidi2021}
E.~Jamshidi, A.~Asgary, N.~Tavakoli, A.~Zali, F.~Dastan, A.~Daaee,
  M.~Badakhshan, H.~Esmaily, S.~H. Jamaldini, S.~Safari, E.~Bastanhagh,
  A.~Maher, A.~Babajani, M.~Mehrazi, M.~A.~S. Kashi, M.~Jamshidi, M.~H.
  Sendani, S.~J. Rahi, N.~Mansouri, Symptom prediction and mortality risk
  calculation for covid-19 using machine learning, medRxiv\href
  {http://arxiv.org/abs/https://www.medrxiv.org/content/early/2021/02/06/2021.02.04.21251143.full.pdf}
  {\path{arXiv:https://www.medrxiv.org/content/early/2021/02/06/2021.02.04.21251143.full.pdf}},
  \href {http://dx.doi.org/10.1101/2021.02.04.21251143}
  {\path{doi:10.1101/2021.02.04.21251143}}.

\bibitem{chowdhury2019comparative}
M.~J.~M. Chowdhury, M.~S. Ferdous, K.~Biswas, N.~Chowdhury, A.~Kayes,
  M.~Alazab, P.~Watters, A comparative analysis of distributed ledger
  technology platforms, IEEE Access 7 (2019) 167930--167943.

\bibitem{FENG201945}
Q.~Feng, D.~He, S.~Zeadally, M.~K. Khan, N.~Kumar,
  \href{https://www.sciencedirect.com/science/article/pii/S1084804518303485}{A
  survey on privacy protection in blockchain system}, Journal of Network and
  Computer Applications 126 (2019) 45--58.
\newblock \href {http://dx.doi.org/https://doi.org/10.1016/j.jnca.2018.10.020}
  {\path{doi:https://doi.org/10.1016/j.jnca.2018.10.020}}.
\newline\urlprefix\url{https://www.sciencedirect.com/science/article/pii/S1084804518303485}

\bibitem{nakamoto2008peer}
S.~Nakamoto, A.~Bitcoin, A peer-to-peer electronic cash system, Bitcoin.--URL:
  https://bitcoin. org/bitcoin. pdf 4.

\bibitem{rinkeby}
rinkeby testnet, \href{https://www.rinkeby.io/#stats}{rinkeby}.
\newline\urlprefix\url{https://www.rinkeby.io/#stats}

\bibitem{chowdhury2013captcha}
M.~J.~M. Chowdhury, N.~R. Chakraborty, Captcha based on human cognitive factor,
  arXiv preprint arXiv:1312.7444.

\bibitem{jmeter}
R.~Abbas, Z.~Sultan, S.~N. Bhatti, Comparative analysis of automated load
  testing tools: Apache jmeter, microsoft visual studio (tfs),loadrunner,
  siege, in: 2017 International Conference on Communication Technologies
  (ComTech), 2017, pp. 39--44.
\newblock \href {http://dx.doi.org/10.1109/COMTECH.2017.8065747}
  {\path{doi:10.1109/COMTECH.2017.8065747}}.

\bibitem{nevedrov2006using}
D.~Nevedrov, Using jmeter to performance test web services, Published on
  dev2dev (2006) 1--11.

\end{thebibliography}

\end{document}